\documentstyle[12pt,axodraw,epsf]{article}
\begin{document}
\baselineskip 0.60cm
\newcommand{\gsim}{ \mathop{}_{\textstyle \sim}^{\textstyle >} }
\newcommand{\lsim}{ \mathop{}_{\textstyle \sim}^{\textstyle <} }
\newcommand{\vev}[1]{ \left\langle {#1} \right\rangle }
\newcommand{\bra}[1]{ \langle {#1} | }
\newcommand{\ket}[1]{ | {#1} \rangle }
\newcommand{\EV}{ {\rm eV} }
\newcommand{\KEV}{ {\rm keV} }
\newcommand{\MEV}{ {\rm MeV} }
\newcommand{\GEV}{ {\rm GeV} }
\newcommand{\TEV}{ {\rm TeV} }
\def\diag{\mathop{\rm diag}\nolimits}
\def\Spin{\mathop{\rm Spin}}
\def\SO{\mathop{\rm SO}}
\def\O{\mathop{\rm O}}
\def\SU{\mathop{\rm SU}}
\def\U{\mathop{\rm U}}
\def\Sp{\mathop{\rm Sp}}
\def\SL{\mathop{\rm SL}}

%%%%%%%%%%
%%%%%%%%%%      title page
%%%%%%%%%%

\begin{titlepage}

\begin{flushright}
UT-959\\
\end{flushright}

\vskip 2cm
\begin{center}
{\large \bf  Semi-Simple Unification on ${\bf T}^6/{\bf Z}_{12}$
 Orientifold \\}
{\large \bf in the Type IIB Supergravity}

\vskip 1.2cm
T.~Watari$^a$ and T.~Yanagida$^{a,b}$

\vskip 0.4cm
$^{a}$ {\it Department of Physics, University of Tokyo, \\
         Tokyo 113-0033, Japan}\\
$^{b}$ {\it Research Center for the Early Universe, University of Tokyo,\\
         Tokyo 113-0033, Japan}

\vskip 1.5cm
\abstract{The semi-simple unification model based on $\SU(5)_{\rm GUT} 
\times \U(3)_{\rm H}$ gauge group is an interesting extension of 
the minimal $\SU(5)_{\rm GUT}$ grand unification theory (GUT), 
since it solves the two serious problems in the standard GUT:
the triplet-doublet splitting problem and the presence of dangerous 
dimension five operators for proton decay.
Here, the extra $\U(3)_{\rm H}$ gauge interaction plays a crucial role on the
 GUT breaking.
In this paper, we show that the full multiplet structure of the 
$\U(3)_{\rm H}$ sector required for the desired GUT breaking is reproduced
 naturally on ${\bf T}^6/{\bf Z}_{12}$ orientifold in the type IIB
 supergravity with a D3-D7 system. 
The $\SU(5)_{\rm GUT}$ vector multiplet lives on D7-branes and 
the  $\U(3)_{\rm H}$ sector resides on D3-branes.
We also show that various interesting features in the original
 $\SU(5)_{\rm GUT} \times \U(3)_{\rm H}$ model are explained in the
 present brane-world scenario.
A possible extension to the type IIB string theory is also discussed.}

\vskip 2cm
PACS code(s): 12.10.-g, 11.25.Mj, 12.60.Jv
\end{center}
\end{titlepage}

%%%%%%%%%%
%%%%%%%%%%      main part 
%%%%%%%%%%

\section{Introduction}

Supersymmetric grand unification theory (SUSY-GUT) is strongly 
supported by the success of the gauge-coupling unification\cite{LL}. 
The minimal SU(5) GUT model, however, suffers from two serious problems; 
one is the triplet-doublet splitting problem and the other is the
presence of the dimension five operators\cite{dim5} causing a too fast
proton decay. Semi-simple unification model based on 
SU(5)$_{\rm GUT}\times$U(3)$_{\rm H}$ gauge group\cite{ss,IY-ss} is an 
extension of the minimal SU(5)$_{\rm GUT}$ model that solves the above 
two problems keeping the success of the original GUT model. In this model 
color-triplet Higgs multiplets acquire large masses of order of the GUT scale
together with their partners, while weak-doublet Higgs multiplets remain 
massless\cite{mpm}. The masslessness of the Higgs doublets 
is guaranteed by an R symmetry and hence the Higgs doublets get
SUSY-breaking scale masses through the Giudice-Masiero
mechanism\cite{GM}.

In a recent article\cite{IWY} Imamura and the present authors have
pointed out that the above semi-simple unification model is quite
naturally embedded in the type IIB supergravity with a D3-D7 brane
system. In this higher dimensional theory various 
phenomenologically important features of the original  semi-simple 
unification model are well understood by the brane-world structure.
For instance, the hierarchy between the GUT scale and the Planck scale 
and the disparity between gauge coupling constants of 
SU(5)$_{\rm GUT}$ and U(3)$_{\rm H}$ are simultaneously explained.
Moreover, a major part of the matter content and interactions that
are most relevant to the GUT breaking is also reproduced from the
D3-D3 and the D3-D7 sector fields.

The purpose of this paper is to show that problems left unsolved in
the previous article\cite{IWY} that are related to the hypercolor U(3)$_{\rm
H}$ sector (relevant to the GUT breaking dynamics) is completely resolved
when one adopts the ${\bf T}^6/{\bf Z}_{12}$ orientifold in the type IIB 
supergravity.

We take a bottom-up approach to construct a model in the ten
dimensional space-time supergravity. We first assume a torus
compactification ${\bf T^6}$ of the extra six dimensional space and
consider the type IIB supergravity with a D3-D7 brane system.
The presence of the D-branes requires that our manifold is no longer 
the torus  ${\bf T^6}$ but ${\bf T}^6/({\bf Z}_2\vev{R_{4567}} 
\times {\bf Z}_2\vev{\Omega R_{89}})$ (notation is explained later)  
since we suppose that the Ramond-Ramond charge of the D3- and D7-branes are 
canceled by orientifold planes O3 and O7\cite{Polchinski}\footnote{
Orientifold $p$ planes are ($p$ + 1) dimensional objects in string 
theories that have  Ramond-Ramond charge opposite to  that of the 
D$p$-branes\cite{Polchinski}. Although 
their existence is not manifest within the supergravity, we assume
their existence because the D-brane charges must be canceled. We call
${\bf T}^6/(\Gamma \times {\bf Z}_2 \vev{\Omega R_{89}})$ as ${\bf
T}^6/\Gamma$ orientifold in this paper.}. As for possible gauge 
symmetries and massless matter multiplets on the D3-D7 system 
we assume those predicted by the type IIB
string theory, since it is known that the string theory provides a 
consistent field theory of massless fields in the ten dimensional 
supergravity. Then, we take the standard procedure of orbifolding to
obtain an N=1 SUSY four dimensional theory. We see that some anomalies 
appear at orbifold fixed points through the orbifold projection. 
Therefore, we introduce new fields at the fixed points to cancel the 
anomalies and obtain a consistent field theory on an orientifold
background in the ten dimensional space-time.
This strategy is very similar to the construction of the Heterotic M theory
on $S^1/{\bf Z}_2$ orbifold in the eleven dimensional
supergravity\cite{hetM}.  

It is very surprising that the complete multiplet in the hypercolor 
U(3)$_{\rm H}$ sector required for the successful phenomenology is
obtained on the ${\bf T}^6/{\bf Z}_{12}$ orientifold. We also find that 
the above mentioned anomalies appear only at a unique ${\bf Z}_{12}$ 
fixed point and they are easily removed by introducing new fields at the 
fixed point. A possible connection to the
type IIB string theory is discussed in the last section.

\section{Brief review of the semi-simple unification model in the brane world}

Let us first review briefly on the semi-simple unification model \cite{IY-ss}.
The gauge group is $\SU(5)_{\rm GUT} \times \U(3)_{\rm H}$. 
Quark, lepton and Higgs supermultiplets are singlets under 
the $\U(3)_{\rm H}$ and transform under the $\SU(5)_{\rm GUT}$ 
as in the standard $\SU(5) _{\rm GUT}$ model. 
Fields introduced for GUT breaking are given as follows:
$X^{\alpha}_{\;\;\beta}(\alpha ,\beta =1,2,3)$ transforming 
as (${\bf 1}$,{\bf adj.}=${\bf 8}+{\bf 1}$) under 
the $\SU(5)_{\rm GUT}\times \U(3)_{\rm H}$ gauge 
group, and $Q^{\alpha}_{\;i}(i=1,\cdots,5)+Q^{\alpha}_{\;6}$
and $\bar{Q}^i_{\;\alpha}(i=1,\cdots,5)+\bar{Q}^6_{\;\alpha}$
transforming as (${\bf 5}^*+{\bf 1}$,${\bf 3}$) and 
(${\bf 5+1}$,${\bf 3}^*$).
Indices $\alpha$ and $\beta$ are for the $\U(3)_{\rm H}$ and $i$ 
for the $\SU(5)_{\rm GUT}$.
Superpotential is given by 
\begin{eqnarray}
W &=& \bar{Q}^k_{\;\alpha} X^{\alpha}_{\;\beta}Q^{\beta}_{\;k} 
      -  v^2 X^\alpha_{\;\alpha}    \nonumber \\
  & & + \bar{H}_i \bar{Q}^i_{\;\alpha} Q^{\alpha}_{\;6} 
      + \bar{Q}^6_{\;\alpha} Q^{\alpha}_{\;i}H^i   \label{eq:super}    \\
  & & + y_{\bf 10} {\bf 10} \cdot {\bf 10} \cdot H 
      + y_{\bf 5^*} {\bf 5}^* \cdot {\bf 10} \cdot \bar{H} + \cdots,
    %  + f   {\bf 5}^*_i \frac{\bar{Q}^i_{\;\alpha}Q^{\alpha}_{\;j}}{M_*^2}
    %           {\bf 10}^{jk} \bar{H}_k, 
\nonumber
\end{eqnarray}
where % $M_*$ denotes the cut off scale of the theory, 
$k = 1,\cdots ,6$ and the parameter $v$ is of order of the GUT scale, 
and $y_{\bf 10}$ and $y_{\bf 5^*}$ are
Yukawa coupling constants of the quarks and leptons.
One can see that the above superpotential has ${\bf Z}_4$ R symmetry
with a charge assignment given in Table \ref{tab:Z4-charge}, and this
symmetry forbids the mass term $W = H\bar{H}$.
The bifundamental representation 
$Q^{\alpha}_{\;i}$ and $\bar{Q}^i_{\;\alpha}$ acquire vacuum expectation
values, $\vev{Q^\alpha_{\;\;i}}= v \delta^\alpha_i$ and 
$\langle$$\bar{Q}^i_{\;\;\alpha}$$\rangle$$= v \delta^i_\alpha$, 
because of the first line in Eq.(\ref{eq:super}), and hence the
gauge group $\SU(5)_{\rm GUT}\times \U(3)_{\rm H}$ is broken down to that
of the standard model\cite{IY-ss}\footnote{
Recently, a similar GUT breaking model that uses expectation values of 
bifundamental fields is discussed in \cite{Csaki}.}.
The mass terms of the colored Higgs multiplets arise from 
the second line in Eq.(\ref{eq:super}) in the GUT-breaking vacuum. The
mass terms of the Higgs doublets are still forbidden by the unbroken 
 ${\bf Z}_4$ R symmetry. One can also see that the ${\bf Z}_4$ R
symmetry forbids the dangerous dimension five proton decay operators 
$W = {\bf 10}\cdot{\bf 10}\cdot{\bf 10}\cdot{\bf 5}^*$. 

There are two remarks here. 
First of all, fine structure constants of the $\U(3)_{\rm H} \simeq
\U(1)_{\rm H} \times \SU(3)_{\rm H}$ must be larger than that of the
$\SU(5)_{\rm GUT}$. 
This is because the gauge coupling constants of the standard model are given by
\begin{equation}
 \frac{1}{\alpha_C} = \frac{1}{\alpha_{\rm GUT}} + \frac{1}{\alpha_{\rm 3H}},
\end{equation}
\begin{equation}
 \frac{1}{\alpha_L} =  \frac{1}{\alpha_{\rm GUT}} \quad \quad \quad
\end{equation}
and 
\begin{equation}
\frac{3/5}{\alpha_Y} = \frac{1}{\alpha_{\rm GUT}} + \frac{2/5}{\alpha_{\rm 1H}}
\end{equation} 
at the GUT scale, where $\alpha_C, \alpha_L, \alpha_Y, \alpha_{\rm GUT}, 
\alpha_{3\rm H}$ and $\alpha_{1\rm H}$ are fine structure constants 
of the three standard model gauge groups, $\SU(5)_{\rm GUT}$, 
$\SU(3)_{\rm H}$ and $\U(1)_{\rm H}$, respectively. 
$1/\alpha_{3H}$ and $1/\alpha_{1H} $ must be within a few \% of the 
$1/\alpha_{\rm GUT}$ to reproduce the approximate
unification of  $\alpha_C, \alpha_L$ and $5 \alpha_Y /3$.
Secondly, the cut-off scale of the theory $M_*$ must be lower than the
Planck scale $M_{\rm Planck} \simeq 2.4 \times 10^{18} \GEV$. 
Indeed, the gauge coupling constant of the $\U(1)_{\rm H}$ is already large 
at the GUT scale, as stated above, and it becomes infinity 
below the Planck scale because of its asymptotic non-free nature. 

%% It has been discussed that the string scale, which will be 
%% the cut off scale of the effective field theory, may be lower than the
%% Planck scale in the type I string theory in order to explain 
%% the hierarchy between the Planck scale and the GUT
 
This relatively low cut-off scale $M_* (\simeq 10^{17} \GEV)$ has
motivated us \cite{IWY} to consider a brane-world in a higher
dimensions. In this brane world the Planck scale is merely 
an effective scale and its relatively large value compared with the
cut-off scale is explained by an effect of a slightly large volume of the 
extra dimensions\cite{Witten}. Interesting is that the disparity 
between the gauge 
coupling constants of the  SU(5)$_{\rm GUT}$ and U(3)$_{\rm H}$ is 
also explained if the  SU(5)$_{\rm GUT}$ gauge multiplet propagates 
in a higher dimensional space (bulk) and the hypercolor U(3)$_{\rm H}$ 
gauge multiplets reside on a ``3-brane''\cite{IWY}.

Moreover, there is another reason\cite{IWY} that supports such a
brane-world structure behind the semi-simple unification model.
The $\U(3)_{\rm H}$ sector (GUT breaking sector), 
which consists of a $\U(3)_{\rm H}$ vector multiplet, 
a $\U(3)_{\rm H}$-adjoint chiral multiplet $X^\alpha _\beta$ and 
vector-like chiral multiplets $Q_k,\bar{Q}^k(k=1,\cdots,6)$, 
has a multiplet structure of the N=2 SUSY, and 
the first line of the superpotential Eq.(\ref{eq:super}) 
% , which is characterized as the superpotential comprised 
% only of the GUT breaking sector fields, 
is a part of the ``gauge interaction'' of 
the N=2 SUSY gauge theory\footnote{N=2 SUSY 
insists the Yukawa coupling of $\bar{Q}XQ$ to be given by the $\U(3)_{\rm
H}$ gauge coupling constant as $\sqrt{2}g_{\rm H}$. 
This property is, however, not necessarily required from phenomenology.} 
with Fayet-Iliopoulos (FI) F term\cite{N=2super}.
Therefore, it seems quite natural to consider that there exists a higher 
dimensional structure as an origin of such an extended SUSY.

In the previous work\cite{IWY} we have found that the type IIB D3-D7 
system on an orientifold geometry is a suitable framework 
to accommodate 
the original semi-simple unification model. The $\U(3)_{\rm H}$ gauge 
symmetry is present on the D3-branes and the $\SU(5)_{\rm GUT}$ 
on the D7-branes. The volume of the D7-branes transverse to the D3-branes
is considered to be larger than the $(1/M_*)^4$ 
to realize the desired features explained above.
%% so that the two features are explained as above.
%%%%%%%%%%%%
%%%%%%%     previous version says.....
%% Perturbative $\SU(5)_{\rm GUT}$ coupling and the weak Newton constant 
%% are explained if the volume of the D7-branes transverse to the D3-branes
%% is larger than the $(1/M_*)^4$.
%%%%%%%%%%%%%
Furthermore, we have considered that the D3-branes on which the 
$\U(3)_{\rm H}$ sector resides are not located at a fixed point, and hence 
in this case the multiplet structure of the N=2 SUSY and the form 
of the N=2 like superpotential in Eq.(\ref{eq:super}) 
are naturally accounted for
by the N=2 SUSY of the D3-D7 system. 
Here, the $\U(3)_{\rm H}$-adjoint $X^\alpha_{\;\beta}$ field arises from 
the D3-D3 sector fields and the bifundamental 
$Q^\alpha_{\;i}, \bar{Q}^i_{\;\alpha}$ ($i=1,\cdots,5$) fields 
from the D3-D7 sector fields. 
%Quarks and leptons are supposed to appear at an orbifold fixed point 
%with only N=1 SUSY.

Although the above D3-D7 system on an orientifold is a good
framework for the semi-simple unification model, 
there are several problems left unsolved.
First of all, 
%it is not clearly described how 
%the U(3)$_{\rm H}\times $U(5)$_{\rm GUT}$ gauge group really 
%appears on the D3-D7 system located on the O7-plane.
%Secondly, 
the origin of the hypermultiplets $Q_6$ and $\bar{Q}^6$ was not clearly found. 
Secondly, the position of the $\U(3)_{\rm H}$ D3-branes in
D7-tangential directions  must be fixed by some dynamics,
or otherwise Nambu-Goldstone (NG) modes of the brane
position remain light and alter the renormalization group (RG)
running of the standard-model gauge coupling constants.
%% Finally!!!, although the origin of the FI F term was
%% discussed in \cite{IWY}, it relies heavily on unknown dynamics.
We show in this paper that all these problems are solved in the 
${\bf T}^6/{\bf Z}_{12}$ orientifold model.
However, the origin of the quarks and leptons is still unclear, 
and hence we put these fields at a four dimensional fixed point by hand.

\section{Model construction on the ${\bf T}^6/{\bf Z}_{12}$ orientifold}

As mentioned in the introduction we adopt the type IIB supergravity with 
a D3-D7 system and assume, on D-branes, gauge symmetries and massless fields 
 obtained from the type IIB string theory, since 
it is known to provide a consistent higher dimensional supergravity. 
%% Thus, we use, in this section, the string-theory construction 
%% to find out a consistent supergravity field theory.

\subsection{Whole $\U(3)_{\rm H}$ sector out of orbifold projection}

N=2 SUSY is preserved in a D3-D7 system on an orientifold 
${\bf T}^6/({\bf Z}_2\vev{R_{4567}} \times {\bf Z}_2\vev{\Omega
R_{89}})$. 
Here, ${\bf Z}_2\vev{R_{4567}}$ is a ${\bf Z}_2$ symmetry
generated by a space reflection in the 4th - 7th 
directions $R_{4567}$, and ${\bf Z}_2\vev{\Omega R_{89}}$ a ${\bf
Z}_2$ symmetry generated by a space reflection $R_{89}$ 
in the 8th and the 9th directions along with an exchange $\Omega$ 
of two Chan-Paton indices.

In the type IIB string theory with a D3-D7 system and 
 O3- and O7-planes the maximal gauge group is 
$\U(16) \times \U(16)$\cite{GP}. 
Each $\U(16)$ is realized on thirty two D3-branes and 
thirty two D7-branes, respectively. 
Gauge group becomes smaller if the D-branes cluster in several places.
The gauge group on D7-branes can be $\prod_{i=1}^4 \U(n_i) \times 
\prod_j \U(n_j)$ ($\sum n_i + \sum n_j =16 $).
The former $\prod_{i=1}^4 \U(n_i)$ come from D7-branes that are fixed
under the $R_{89}$ %({\it i.e.,} that are located on the four O7-planes) 
and the latter $\prod_j \U(n_j)$ come from 
the rest of the D7-branes, which are not fixed under the $R_{89}$ 
({\it i.e.,} that are located away from the O7-planes).
The gauge group on D3-branes can be $\prod_{i=1}^{64} \U(m_i) \times 
\prod_j Sp(2 m_j) \times \prod_k \U(m_k) \times \prod_l \U(m_l)$ 
($\sum m_i +2 \sum m_j + \sum m_k + 2 \sum m_l = 16$).
The $\prod_{i=1}^{64} \U(m_i)$ come from D3-branes fixed both under the
$R_{4567}$ and the $R_{89}$, the $\prod_j \Sp(2 m_j)$ from those fixed 
only under the $R_{89}$, the $\prod_m \U(m_k)$ from those fixed only
under the $R_{4567}$ and the $\prod_l \U(m_l)$ from the rest of the
D3-branes% ({\it i.e.,} those are fixed neither under 
% the $R_{89}$ nor $R_{4567}$). 
\cite{GP}. Gauge group can easily be $\U(n) \times \U(m) \times \cdots $.
Therefore, it is quite natural to incorporate $\SU(5)_{\rm 
GUT} \times \U(3)_{\rm H} \times \cdots $ gauge group in this framework.
%% In this paper, we identify $\SU(5)_{\rm GUT}$ with a subgroup of
%% $\U(n_j)$ and 
%% $\U(3)_{\rm H}$ with some of $\U(m_k)$ as one sees in the following.

% We take this gauge theory on the D3-D7 system with the matter content
% predicted from the type IIB string theory at the starting point of the
% model construction. 
% Such a matter content would not cause any 
% inconsistencies of the higher dimensional field theories. 

Now let us consider the orbifolding of the 
${\bf T}^6/({\bf Z}_2\vev{R_{4567}} \times {\bf Z}_2\vev{\Omega
R_{89}})$ to reduce the N=2 SUSY down to the N=1 SUSY.
% We take twelve D7-branes and we have $\U(6)$ gauge group 
% as the origin of the
% $\SU(5)_{\rm GUT}$. Remaining D7-branes are expected to be away from the
% D7-branes required here. Since the $\SU(5)_{\rm GUT}$ charged 
% quarks and leptons are chiral and are at most N=1 supersymmetric, 
% we impose orbifold projection on the $\U(6)$ gauge theory. 
We assume ${\bf Z}_N$ symmetry of the six dimensional torus ${\bf
T}^6$ as a candidate of the orbifold group. 
In order to preserve the N=1 SUSY, the ${\bf Z}_N$ rotational symmetry, 
which belongs to the six dimensional rotational group $\SO(6)\simeq \SU(4)$, 
must be included in $\SU(3)\subset\SU(4)$ that rotates three
complex planes holomorphically\cite{orbifoldII,notation}. 
Thirteen candidates that preserve N=1 SUSY are listed in \cite{orbifoldII}. 
Among these, ${\bf Z}_N$ that contains ${\bf Z}_2\vev{R_{4567}}$ as a
subgroup and that breaks the N=2 down to the N=1 SUSY is desirable 
for our purpose. 
In this paper, we use ${\bf Z}_{12}\vev{\sigma}$ where 
${\bf Z}_2\vev{R_{4567}}\simeq {\bf Z}_2\vev{\sigma^6} \subset 
{\bf Z}_{12}\vev{\sigma}$; 
namely we consider ${\bf T}^6/{\bf Z}_{12}$ orientifold model ({\it i.e.,}
${\bf T}^6/({\bf Z}_{12}\vev{\sigma} \times {\bf Z}_2\vev{\Omega
R_{89}})$)\footnote{There are two ${\bf Z}_{12}$ candidates. 
The ${\bf Z}_{12}$ we use in this paper is the one written as ${\bf Z}_{12}$ in
\cite{notation} and 
not the one written as ${\bf Z}_{12}'$ in \cite{notation}. We adopt the 
notation of the orbifold group in \cite{notation}.}. 
Other possibilities are discussed in a future 
publication\cite{future}.

% Thus, the orientifold geometry must have 7+1 dimensional orbifold 
% fixed loci away from the O7-planes. Under this constraint we have only
% three candidates ${\bf Z}_6$, ${\bf Z}_6'$ and ${\bf Z}_{12}$ described in
% \cite{orbifoldII}. In the following discussion we are restricted in a 
% case of 

% Here, $\sigma$ denotes the generator of the ${\bf Z}_{12}$ symmetry of
% the six dimensional torus.  
There are three 7+1 dimensional ${\bf Z}_{12}$ fixed loci, among which 
one coincides with an O7-plane and the other two are mirror images 
of each other under the ${\Omega R_{89}}$ (see Fig.\ref{fig:89}). 
We put a pair of six+six D7-branes on this mirror pair of fixed loci 
on which a $\U(6)$ gauge group is obtained. The reason why we do not 
choose the one on the O7-plane but the latter 
fixed loci will be explained later.
Before operating the ${\bf Z}_{12}$ orbifold projection  
the gauge theory on the D7-branes consists of an N=4 $\U(6)$ 
vector multiplet---an N=1 vector multiplet $\Sigma_{(0)}\equiv 
{\cal W}_\alpha^{\U(6)}$ and three N=1 chiral multiplets $\Sigma_{(b)}$
(b=1,2,3)---, which is, however, subject to the the ${\bf Z}_{12}$ 
orbifold projection, since the D7-branes are fixed under the ${\bf Z}_{12}$. 
This $\U(6)$ group is the origin of the $\SU(5)_{\rm GUT}$, 
as derived in the following. 
Remaining twenty D7-branes are supposed to be away from these twelve 
D7-branes. 
% ((They might be an origin of hidden sector gauge group.))

We impose a ${\bf Z}_{12}\vev{\sigma}$ orbifold projection  on the
D7-D7 sector fields ($\Sigma_{(a)}$ (a=$0,\cdots,3$)) on the twelve 
D7-branes we are interested in. 
The generator $\sigma$ is given by
\begin{equation}
\sigma = e^{-2\pi i \diag(v_a)_{a=0,\cdots,3}} =
     e^{-2\pi i \diag(0,\frac{1}{12},\frac{-5}{12},\frac{4}{12})} \in 
 \SU(3) \subset \SU(4),
\end{equation}
which rotates three complex planes as
\begin{equation}
 z_b \rightarrow e^{2\pi i v_a} z_b,\quad \quad (b = 1,2,3) 
\label{eq:generator2}
\end{equation}
where $z_b$  denotes b-th complex coordinate of the six
dimensional torus.
%%%%%  The ${\U(6)}$ N=4 multiplet 
%%  with (12 $\times$ 12)-matrix Chan-Paton wave function 
%%  $\lambda_a $ ($a = 0,\cdots,3$)\footnote{Note that the Chan-Paton wave
%%  function $\lambda^a$ are already under a constraint
%%  \begin{equation}
%%  \lambda^a = \left(\begin{array}{cc}
%%  	     & 1\\ 1&
%%  		  \end{array}\right)\lambda^{a T}
%%              \left(\begin{array}{cc}
%%  	     & 1\\ 1&
%%  		  \end{array}\right)		  
%%  \end{equation} 
%%  because of the orientifold projection due to ${\bf Z}_2\vev{\Omega
%%  R_{89}}$. Off diagonal 6 $\times$ 6 blocks of the $\lambda^a$ 
%%  are also irrelevant(massive) because the pair of the six+six D7-branes 
%%  are separated in the 3rd complex plane, as is seen in Fig.\ref{fig:89}.}
%%  on the relevant six+six D7-branes 
The massless spectrum of the gauge theory in the orientifold geometry is 
given by partial components of the $\U(6)$ N=4 multiplet $\Sigma_{(a)}$ 
that satisfy the ${\bf Z}_{12}\vev{\sigma}$ orbifold projection 
condition \cite{notation}:
\begin{equation}
    \Sigma^{\;\;k}_{(a)\;l} = e^{2\pi i v_a} 
         \left(\widetilde{\gamma}_{\sigma,7} 
            \Sigma_{(a)} \widetilde{\gamma}_{\sigma,7}^{-1}\right)^k_{\;\;l},
\label{eq:orb-prj}
\end{equation}
where we take the (6 $\times$ 6) matrix $\widetilde{\gamma}_{\sigma,7}$ as
\begin{equation}
%%  \gamma_{\sigma,7} = \diag(
%%           \overbrace{e^{-\pi i\frac{1}{12}},\cdots,e^{-\pi i\frac{1}{12}}}^5,
%%                      e^{-\pi i \frac{3}{4}},
%%           \overbrace{e^{\pi i\frac{1}{12}},\cdots,e^{\pi i\frac{1}{12}}}^5,
%% 	             e^{\pi i \frac{3}{4}}). 
 \widetilde{\gamma}_{\sigma,7} = \diag(
          \overbrace{e^{-\pi i\frac{1}{12}},\cdots,e^{-\pi i\frac{1}{12}}}^5,
                     e^{-\pi i \frac{3}{4}}).
\label{eq:cp-7-s1}
\end{equation}
Notice that the (12 $\times$ 12) matrix 
$\gamma_{\sigma,7}$ acting on twelve D7-Chan-Paton indices 
that appears frequently in literatures is given by 
\begin{equation}
   \gamma_{\sigma,7} = \diag(\widetilde{\gamma}_{\sigma,7},\widetilde{\gamma}_{\sigma,7}^{-1}).   
\end{equation} 
Eq.(\ref{eq:cp-7-s1}) is chosen so that the surviving gauge group 
is $\U(5) \times \U(1)_6$. The
$\SU(5)$ subgroup of the $\U(5)\times \U(1)_6$ is identified with the 
$\SU(5)_{\rm GUT}$. 
We see easily that only an N=1 chiral multiplet $\Sigma^6_{(3)\;i}$ 
which transforms as $({\bf 5}^*,+1)$ under the gauge group 
$\U(5) \times \U(1)_6$ survives the orbifold projection besides the N=1
vector multiplets from the $\Sigma_{(0)}$.
This multiplet comes from the fluctuation mode of the D7-branes
(a massless mode of D7-D7 open string) in their transverse directions $z_3$. 
Anomaly cancellation conditions and the fate of the two remaining $\U(1)$ 
gauge symmetries are discussed later. 
We will identify the N=1 chiral multiplet $\Sigma^6_{\;\; i}({\bf 5}^*)$ 
with one of the Higgs multiplets, $\bar{H}_i({\bf 5}^*)$, in later arguments.

Now let us discuss how we obtain suitable massless spectrum from
the D3-D3 sector and the D3-D7 sector. 
If the $\U(3)_{\rm H}$ D3-branes are located in the bulk, 
unwanted NG modes destroy the gauge-coupling unification. 
If the D3-branes are located at an orbifold fixed point, however, 
then the N=2 multiplet structure required in the $\U(3)_{\rm H}$ sector 
might be lost.  
One way out of this difficulty is given by noticing 
that nature of all the fixed points is not necessarily the same.

Here, we introduce a notion of ``N=2 fixed point''. 
There is an isotropy subgroup $G_x \subset {\bf Z}_{12}$ for each point 
$x$ of the ${\bf T}^6$; $G_x$ is given by elements 
of the ${\bf Z}_{12}$ that fix the point $x$. If $G_x$ rotates only 
the first two complex planes $z_{1},z_{2}$, or in other words, 
$G_x$ is included in an $\SU(2) \subset \SU(3)$ subgroup that  
rotates only the $z_{1},z_{2}$, then we call such point $x$ as 
an ``N=2 fixed point''. 
%% We note here that the ``N=2 fixed points'' always appear as 5+1 dimensional 
%% loci.

Suppose that the D3-branes are located at an ``N=2 fixed
point''\footnote{There is no N=2 multiplet structure on the D3-branes at 
other fixed points.} $x$.
Then there exist mirror images under the ${\bf Z}_{12}/G_x$. 
Orbifold projection due to the ${\bf Z}_{12}/G_x$ identifies all mirror
images, and the identified D3-D3 and the D3-D7 sectors are subject only
to the remaining orbifold projection of the $G_x$. 
Since the $G_x$ at the D3-brane position belongs to the 
$\SU(2) \subset \SU(3)$, the N=2 multiplet structure 
in the D3-D7 system survive the orbifold projection. 
Moreover, we expect that the unwanted NG modes associated with 
the $\U(3)_{\rm H}$ 
D3-brane positions are eliminated since the D3-branes 
can be ``fixed'' at that ``N=2 fixed point''. 

We put the the $\U(3)_{\rm H}$ D3-branes  
at a point $x$ on the $\SU(5)_{\rm GUT}$ D7-branes where the
isotropy subgroup $G_x$ is ${\bf Z}_4\vev{\sigma^3} \subset 
{\bf Z}_{12}\vev{\sigma}$.
It is easy to see that they are ``N=2 fixed points''.
There are only six such points, which form a coset space 
$({\bf Z}_{12}\vev{\sigma} \times {\bf Z}_2\vev{\Omega R_{89}})
/{\bf Z}_4\vev{\sigma^3}$ (see Eq.(\ref{eq:generator2}));
%% namely the ${\bf Z}_{12}\vev{\sigma} \times {\bf
%% Z}_{2}\vev{\Omega R_{89}}$ acts transitively on the six such points. 
namely, all these six points are mirror images of each other.
Therefore, there is essentially only one candidate for the position of
the $\U(3)_{\rm H}$ D3-branes. %at which we put the  $\U(3)_{\rm H}$ sector.
We put three D3-branes at each mirror image and hence we need eighteen
D3-branes\footnote{We put the rest of the D3-branes (there are twelve
D3-branes left since we use two more D3-branes in our model construction 
later.), for example, on the remaining D7-branes. Those D3- and
D7-branes may be used for the dynamical SUSY breaking.}.
The gauge groups $\U(3) \times \U(3) \times \U(3)$ arising from these
eighteen D3-branes in 
%% ${\bf T}^6/({\bf Z}_2 \vev{R_{4567}}\times {\bf Z}_2\vev{\Omega
%% R_{89}})$ orientifold 
the ${\bf T}^6/{\bf Z}_2 \vev{R_{4567}=\sigma^6}$ orientifold\footnote{
These D3-branes are fixed under the $R_{4567}$, while non-fixed under
the $R_{89}$. This is the reason why the gauge group is
$\U(3) \times \U(3) \times \U(3)$.} are identified
as a single $\U(3)$ after the orbifold projection of the ${\bf
Z}_{12}\vev{\sigma}/{\bf Z}_4\vev{\sigma^3}$. 

After this identification under the ${\bf Z}_{12}\vev{\sigma}/{\bf
Z}_4\vev{\sigma^3}$, the D3-D3 and the D3-D7 sector fields consist of 
a $\U(3)$ N=2 vector multiplet ($X_{(0)}\equiv {\cal
W}_{\alpha}^{\U(3)}$,$X_{(3)}$) and a $\U(6)_{\rm D7} \times \U(3)_{\rm D3}$ 
bifundamental N=2 hypermultiplet ($Q^\alpha_{\;\;k}$
,$\bar{Q}^k_{\;\;\alpha}$)$_{k=1,\cdots,6;\alpha=1,2,3}$. 
The former comes from an open string massless mode that starts and ends 
on the three D3-branes and the latter is an open string that starts from 
three D3-branes and ends to six D7-branes and vice versa.
They receive the orbifold projection of the isotropy group 
${\bf Z}_4\vev{\sigma^3}$ and some of them might disappear from the
spectrum of the theory in the full ${\bf Z}_{12}$ orientifold geometry. 

However, the projection conditions
\begin{equation}
 X_{(0),(3)} = \widetilde{\gamma}_{\sigma^3,3} X_{(0),(3)}
                 \widetilde{\gamma}_{\sigma^3,3}^{-1} ,  
\end{equation}
\begin{equation}
 Q^\alpha_{\;\;k} = e^{\pi i 3 v_3}\left(\widetilde{\gamma}_{\sigma^3,3} 
            Q \widetilde{\gamma}_{\sigma^3,7}^{-1}\right)^{\alpha}_{\;\;k}, 
 \quad \quad 
 \bar{Q}^k_{\;\;\alpha} = e^{\pi i 3 v_3} 
            \left(\widetilde{\gamma}_{\sigma^3,7} 
        \bar{Q} \widetilde{\gamma}_{\sigma^3,3}^{-1}\right)^{k}_{\;\;\alpha}
\end{equation}
remove none of the above N=2 multiplets if the (3 $\times$ 3) 
matrix $\widetilde{\gamma}_{\sigma^3,3}$\footnote{The (18 $\times$ 18)
matrix $\gamma_{\sigma,3}$ acting on eighteen D3 Chan-Paton indices
 that frequently appears in literatures is given by 
\begin{equation}
\gamma_{\sigma,3}=\left(\begin{array}{cccccc}
0&*&0& & & \\  0&0&*'& & & \\ **''&0&0& & & \\ 
 & & &0&0&*^{''-1} \\  & & &*^{-1}&0&0 \\  & & &0&*^{'-1}&0 
		  \end{array}\right)
\end{equation}
with (3 $\times$ 3) matrices $* \cdot *' \cdot *'' = *' \cdot *'' \cdot * 
= *'' \cdot * \cdot *' =\widetilde{\gamma}_{\sigma^3,3}$ so that
\begin{equation}
 \left(\gamma_{\sigma,3}\right)^3 = \gamma_{\sigma^3,3} \equiv \diag(
   \widetilde{\gamma}_{\sigma^3,3},
   \widetilde{\gamma}_{\sigma^3,3},
   \widetilde{\gamma}_{\sigma^3,3},
   \widetilde{\gamma}_{\sigma^3,3}^{-1},
   \widetilde{\gamma}_{\sigma^3,3}^{-1},
   \widetilde{\gamma}_{\sigma^3,3}^{-1},).
\label{eq:cp-3-s1-1}
\end{equation}} 
is taken as
\begin{equation}
 \widetilde{\gamma}_{\sigma^3,3} = \diag(
       e^{-\frac{5}{4}\pi i},e^{-\frac{5}{4}\pi i},e^{-\frac{5}{4}\pi i}).
\label{eq:cp-3-s3-1}
\end{equation}
Here, we note that 
\begin{equation}
 \widetilde{\gamma}_{\sigma^3,7} = 
 \left(\widetilde{\gamma}_{\sigma,7}\right)^3.
\end{equation}
%%  We take the (18 $\times$ 18) matrix $\gamma_{\sigma^3,3}$ that are 
%%  used in the orbifold projection on the D3-brane Chan-Paton index 
%%  % by the generator $\sigma^3$, as in Eq.(\ref{eq:orb-prj}), 
%%  is given by
%%  \begin{equation}
%%  \gamma_{\sigma^3,3} = \diag(
%%         \overbrace{
%%            \overbrace{e^{-\frac{5}{4}\pi i},\cdots,e^{-\frac{5}{4}\pi i}}^3 
%%                   ,\cdots,\cdots,}^3
%%         \overbrace{
%%            \overbrace{e^{\frac{5}{4}\pi i},\cdots,e^{\frac{5}{4}\pi i}}^3 
%%                   ,\cdots,\cdots}^3       ). 
%%  \end{equation}
%%  The (12 $\times$ 12) matrix $\gamma_{\sigma^3,7}$ used on the D7-brane 
%%  Chan-Paton index % associated to the generator $\sigma^3$ 
%%  is calculated from Eq.(\ref{eq:cp-7-s1}) as
%%  \begin{equation}
%%  \gamma_{\sigma^3,7} = \diag(
%%         \overbrace{e^{-\frac{1}{4}\pi i},\cdots,e^{-\frac{1}{4}\pi i}}^5,
%%                    e^{-\frac{1}{4}\pi i},
%%         \overbrace{e^{\frac{1}{4}\pi i},\cdots,e^{\frac{1}{4}\pi i}}^5,
%%                    e^{\frac{1}{4}\pi i}  ).      
%%  \end{equation}
%%  Then we can see that the matter spectrum out of the orbifold projection 
%%  is $X^{\alpha}_{\;\;\beta}$ from the D3-D3 sector and
%%  $Q^\alpha_{\;k},\bar{Q}^{k}_{\;\;\alpha}$($k=1,\cdots,6$) from the D3-D7
%%  sector. 
We identify the $\U(3)_{\rm D3}$ gauge group with the $\U(3)_{\rm H}$ and 
we see that the phenomenologically required N=2 multiplets 
of the $\U(3)_{\rm H}$ sector are fully obtained at this ``N=2 fixed points''. 
D3-branes are really fixed at that ``N=2 fixed point'' and there is 
no unwanted massless field that destroys the gauge-coupling unification. 
We show in \cite{future} that similar  arguments are possible in the case of 
${\bf Z}_6$ and ${\bf Z}_6'$ orbifolding.

Unwanted matter multiplets would actually arise if $\Omega
R_{89}$-mirror images of the D3-branes were not separated from themselves,
and that is why we put the D7-branes for the $\SU(5)_{\rm GUT} \subset
\U(6)$ (and of course D3-branes, too) away from the 
O7-planes\footnote{This is also the reason why we do not choose 
${\bf Z}_4$, ${\bf Z}_8$, ${\bf Z}_8'$ or ${\bf Z}_{12}'$ as an orbifold 
group. They have no 7+1 dimensional fixed locus which does not coincide
with the O7-planes.}.

\subsection{Triangle anomaly cancellation}

%% Higher dimensional field theories are subject to much more stringent
%% consistency conditions than those of the four dimensional ones.
%% In order to satisfy these stringent conditions, we used the massless 
%% field content of the type IIB string theory---D3-D7 system on type IIB
%% supergravity. Matter content of the orbifolded geometry is derived from
%% the above matter content through the ordinary procedure of the
%% orbifolding, or otherwise various inconsistency would appear as is
%% discussed in \cite{GP}. We restrict ourselves that we introduce more
%% particles by hand if necessary only on four dimensional N=1 fixed
%% points. 
%% We expect that such introduction of new particles by hand in such a way 
%% would not lead to an inconsistency of higher dimensional field theories
%% and is subject only under usual consistency conditions of ordinary four 
%% dimensional field theories.
%% Detailed discussion will be given in a future publication\cite{future}.

Since we have started our model construction based on the matter content
predicted by the type IIB string theory, there is no inconsistency 
before the ${\bf Z}_{12}$ orbifold projection.
Inconsistencies appear only through the orbifolding
process, and they are expected to localize at orbifold fixed
points\cite{CH,Witten}. 
Indeed, for example, the matter content derived so far has four dimensional
gauge anomalies\footnote{There is another model in which there is no
triangle anomaly by the matter content derived from the
D-branes. There, we put D3-branes at a point on 
the D7-branes where the
isotropy subgroup is ${\bf Z}_2\vev{\sigma^6}$. It is possible in this
model to choose $\gamma_{\sigma,7}$ and $\gamma_{\sigma,3}$ so that no
N=1 chiral multiplet from the D7-D7 sector survives the orbifold
projection. Gauge group is $\SU(5)_{\rm GUT} \times \U(2)_{\rm H}$
instead of the $\SU(5)_{\rm GUT} \times \U(3)_{\rm H}$ and the
weak-doublet Higgs multiplets are identified with the composites 
$\bar{Q}^6_{\alpha}Q^\alpha_{i}$ and $Q^\alpha_6\bar{Q}_{\alpha}^i$. 
See \cite{future} for details.}, 
and a calculation of gauge triangle anomalies like that
given in \cite{anomaly-orbifold} shows that the anomalies localize only 
at four dimensional N=1 fixed points (see \cite{future} for a detailed 
calculation).
% in all ${\bf Z}_6$, ${\bf Z}_6'$ and ${\bf Z}_{12}$ model.

The cancellation condition of these anomalies is stronger than 
that in the ordinary four dimensional field theories. Not the total sum
of the anomalies from all fixed points but also the anomalies at each 
fixed point must be canceled. 
% Among the three models ${\bf Z}_6$, ${\bf Z}_6'$ and ${\bf Z}_{12}$, 
The ${\bf Z}_{12}$ model is special in that all triangle 
anomalies appear only at a single fixed point on the $\U(6)$ D7-branes,
that is, the ${\bf Z}_{12}\vev{\sigma}$ fixed point\cite{future}. 
%% ((We discuss the anomaly cancellation of % the  ${\bf Z}_{12}$ 
%% this model hereafter.)) 
%% Possibility of ${\bf Z}_6$ and ${\bf Z}_6'$ models are discussed 
%% in \cite{future}.

We put first one D3-brane at the ${\bf Z}_{12}\vev{\sigma}$ fixed
point (along with its O7-mirror image% under $\Omega R_{89}$
) to cancel the $\SU(5)_{\rm GUT}$ triangle anomaly. 
The D3-D7 matters must satisfy the orbifold projection condition
\begin{equation}
 \Psi_k = e^{\pi i v_3} \left(\widetilde{\gamma}'_{\sigma,3}\Psi 
                              \widetilde{\gamma}_{\sigma,7}^{-1}\right)_k, 
 \quad \quad 
 \bar{\Psi}^k = e^{\pi i v_3} \left( \widetilde{\gamma}_{\sigma,7}\bar{\Psi}
                               \widetilde{\gamma}_{\sigma,3}^{'-1}\right)^k
\end{equation}
where $k=1,\cdots,6$.
We can easily see that only one N=1 SUSY chiral multiplet
$\bar{\Psi}^i$(${\bf 5}$,0) ($i=1,\cdots,5$ ) of the D7-D7 
$\U(5)\times \U(1)_6$ gauge group satisfies this condition when\footnote{
The (20 $\times$ 20) matrix $\gamma_{\sigma,3}$ is given by (18 $\times$ 18
part Eq.(\ref{eq:cp-3-s1-1})) $\oplus$ 
diag($\widetilde{\gamma}'_{\sigma,3},\widetilde{\gamma}^{'-1}_{\sigma,3}$).}
\begin{equation}
\widetilde{\gamma}'_{\sigma,3} = e^{\frac{1}{4}\pi i}.
\label{eq:cp-3-s1-2}
\end{equation}
The $\SU(5)_{\rm GUT}$ triangle anomaly is completely canceled by this 
$\bar{\Psi}^i({\bf 5})$. 
We can identify the multiplet $\bar{\Psi}^i({\bf 5})$ with the 
Higgs multiplet 
$H^i({\bf 5})$.
D3-D3 sector provides another (anomalous) $\U(1)_X$ N=1 vector 
multiplet\footnote{This pair of two D3-branes are put at a pair of points,
which is fixed by $R_{4567}$ and not by $R_{89}$. Thus the gauge symmetry
is $\U(1)$.}.
Now there are three $\U(1)$ gauge symmetries: $\U(1)_5\equiv 
({\rm the~centre~of}\;\U(5))$, $\U(1)_6$ and $\U(1)_X$.
Since the matter $\Sigma^6_{(3)\;i}({\bf 5}^*)$ and the
$\bar{\Psi}^i({\bf 5})$ 
transform as $(-1,1,0)$ and $(1,0,-1)$ under
these three $\U(1)$ symmetries, respectively, only a linear combination
$\U(1)_6-\U(1)_X$ has mixed anomalies $\U(1)\cdot [\SU(5)]^2$ and
$\U(1)\cdot [{\rm grav.}]^2$. These anomalies are canceled by
an introduction of a field on the ${\bf Z}_{12}$ fixed point 
that shifts under this $\U(1)$ gauge symmetry\footnote{Such an 
introduction of matter into a four dimensional fixed point with
only N=1 SUSY would not lead to an inconsistency of higher dimensional
field theories.}\cite{genGS}. Each of the remaining two $\U(1)$ symmetries 
can be identified with the $\U(1)_{\rm B-L}$ symmetry. 
Triangle anomaly of this U(1) symmetry vanishes. 
Mixed anomalies between the $\U(1)_{6-X}$ and the $\U(1)_{\rm B-L}$ 
can be canceled by the shifting field introduced above. 
Thus, all triangle anomalies are canceled out.

We introduce the quarks and leptons $3 \times ({\bf 5}^* +{\bf 10})$ 
also at this ${\bf Z}_{12}\vev{\sigma}$ fixed point. Although the
$\U(1)_{\rm B-L}$ symmetry has a triangle anomaly, this anomaly is
canceled by three families of right-handed neutrinos.

\section{Phenomenology}

%% superpotential and R symmetry
We identify the N=1 chiral multiplet $\Sigma^6_{(3)\;i}$ ((${\bf
5}^*$,+1) under the $\U(5) \times \U(1)_6$) from the D7-D7 sector 
with $\bar{H}_i$. % , and the (${\bf 5}$,0) multiplet from the D3-D7 sector 
% at the ${\bf Z}_{12}$ fixed point with $H({\bf 5})$.
Since the origin of the $\Sigma^6_{(3)\;i}$ is the fluctuation
of the six D7-branes in their transverse ($z_3$) directions, 
% there exists superpotential 
interactions in the D3-D7 system give rise to a superpotential
\begin{equation}
 W = \sqrt{2}g_{\rm GUT}
    Q^\alpha_{\;\; 6} \Sigma^6_{(3)\; i} \bar{Q}^i_{\;\;\alpha},
\label{eq:mpm-d}
\end{equation}
along with the ``N=2 gauge interaction''
\begin{equation}
 W = \sqrt{2}g_{\rm H}
     \bar{Q}^k_{\;\alpha} X^{\;\alpha}_{(3)\;\beta} Q^\beta_{\;\;k}.
\end{equation} 
Eq.(\ref{eq:mpm-d}) automatically provides the first term in the second
line of the Eq.(\ref{eq:super}).

We have to remember that in supersymmetric higher dimensional theories R
symmetry has its geometrical interpretation. 
R symmetries arise from the local Lorentz symmetry and the
transformation property under the rotational symmetry determines the R
charge of each field.
Symmetry of the ${\bf T}^6/({\bf Z}_{12} \times 
{\bf Z}_2\vev{\Omega R_{89}})$ geometry contains ${\bf Z}_{4}$ 
R subgroup that rotates the third complex plane by angle $\pi$, 
% as in \cite{IWY}, 
since % now 
the $\Omega R_{89}$ symmetry connects
$z_3$ with $-z_3$. 
Charge assignment under this rotation for the particles derived
from the D3-D3, D3-D7 and D7-D7 sectors is given by $+2$ for 
$X^\alpha_{\;\;\beta}=X^{\alpha}_{(3)\beta}$, $0$ for
$Q^{\alpha}_k$ and $\bar{Q}^k_{\;\alpha}(k=1,\cdots,6)$, and $+2$ for 
$\bar{H}_i=\Sigma^{6}_{(3)i}$. Low energy ${\bf Z}_4$ R symmetry is
considered to be a linear combination of such rotational symmetry and
some $\U(1)$ gauge symmetries. 
Now we have $\U(1)_5$ and $\U(1)_{6+X}$ symmetry. 
It is an easy task to determine the contribution of these two 
$\U(1)$ symmetries so that the charges of 
the $Q_i,\bar{Q}^i (i=1,\cdots,5)$ and $Q_6,\bar{Q}^6$ are those 
given in Table \ref{tab:Z4-charge}. 
Then, we find that the ${\bf Z}_4$ R charge of the $\bar{H}$ becomes 
automatically $0$, the charge given in Table \ref{tab:Z4-charge}.
This nontrivial fact is not coincident, but rather inevitable,
since any relevant symmetry cannot forbid the missing partner mass 
term Eq.(\ref{eq:mpm-d}). 

Another N=1 chiral multiplet $\bar{\Psi}^i({\bf 5})$, which comes 
from the D3-D7
sector at the ${\bf Z}_{12}$ fixed point, can be identified with the
$H({\bf 5})^i$ because of its gauge charge.
Although $H({\bf 5})^i$ and $Q^\alpha_{\;i},\bar{Q}^6_{\;\alpha}$
localize at different fixed points, we expect that exchange of 
particles 
% that are Kaluza-Klein particles of the D7-D7 sector fields or bulk
% particles 
whose masses are of order of the cut-off scale $M_*$ provides the 
superpotential $\bar{Q}^6_{\;\;\alpha}Q^{\alpha}_{\;\;i}H({\bf 5})^i$. 
% If a Kaluza-Klein particle do the job, then there is no problem. 
% Even in the case where the superpotential is provided only
% through 
The exchange of such particles % with $M_*$ mass 
has a suppression factor due to the Yukawa damping of the wave function.
However, the suppression factor is of order $0.1$, as
we see below.% and we can expect that missing partner term.  

%%% neumenology 1  volume and e^-(ML)
In order to explain the disparity of the two gauge coupling constants,
as described in section 2, 
the volume in which the D7-D7 sector fields propagate must be larger 
than the order $M_*^{-4}$.
%% The six dimensional torus ${\bf T}^6$ of the ${\bf Z}_{12}$ model has
%% two basis vectors within the third complex plane $z_3$ (8th and 9th 
%% directions) and four basis vectors in the first $z_1$ and the second
%% $z_2$ complex planes---this four dimensional torus cannot be factorized 
%% into two two-dimensional tori within first and second complex plane, 
%% respectively. 
%% Thus, the 4th - 7th directions should have larger dimensions compared with the
%% fundamental length $M_*^{-1}$.
The four dimensional volume in the 4th -7th directions is $3L^4$
when the distance between the ${\bf Z}_{12}\vev{\sigma}$ fixed point and 
the ${\bf Z}_4\vev{\sigma^3}$ fixed point is given by $L$. 
Gauge-coupling unification condition
\begin{equation}
 \left( \frac{1}{\alpha_{3H}},\frac{1}{\alpha_{1H}} \sim \frac{1}{\alpha_*} 
 \right)
  \lsim 
  10^{-2} \left(\frac{1}{\alpha_{\rm GUT}} \sim \frac{3 (LM_*)^4}{\alpha_*}\right)
\end{equation}
determines the length of four dimensional torus in the 4th - 7th
directions\footnote{Gravity also propagates in the 4th - 7th dimensional
bulk and the Planck scale is determined as $M_P^2 \sim 3(LM_*)^4
M_*^2$ provided that the volume of the 8th and 9th directions is of
order $M_*^{-2}$. The factor $3(LM_*)^4$ determined from the 
gauge-coupling unification condition ($(M_*L)^4 \simeq 30$) 
is large enough to guarantee the low cut-off scale $M_* \sim 10^{17} \GEV$.}. 
In particular, we find $e^{-M_*L} \simeq 0.1$ for $(M_*L)^4 \simeq 30$, 
which is the Yukawa damping of 
wave functions due to the separation between the ${\bf
Z}_{12}\vev{\sigma}$-  and ${\bf Z}_4\vev{\sigma^3}$-fixed point for a
particle of mass $M_*$. 

%%% neumenology 2  small tan beta
In this model, $\bar{H}_i({\bf 5}^*)$ propagates in a bit large extra
dimensions while $H^i({\bf 5})$ does not. 
%% This characteristic feature
%% leads to an interesting phenomenological consequence qualitatively.
%% The kinetic term of the $H$ and Yukawa couplings of the quarks and
%% leptons are given in a four dimensional Lagrangian
%% \begin{eqnarray}
%%  \int d^4x d^4\theta \left(K=H^\dagger H \right)   \\
%%  \int d^4x d^2\theta \left(W = y_* H {\bf 10} {\bf 10} 
%%                              + y_* \bar{H} {\bf 10} {\bf 5}^*
%% \right).
%% \end{eqnarray} 
%% On the other hand, the kinetic term of the $\bar{H}$ is written in eight 
%% dimensional integration
%% \begin{equation}
%%  M_*^4 \int d^4x d^2z_1 d^2z_2 d^4\theta \left( K= \bar{H}^\dagger \bar{H}
%% (x,z_1,z_2) \right),
%% \end{equation}
%% which gives rise to a large wave function renormalization 
%% \begin{equation}
%%  3(M_*L)^4 \int d^4x d^4\theta \left( K= \bar{H}^\dagger \bar{H}(x) \right) 
%% \end{equation}
%% after dimensional reduction to the four dimensions. This large wave
%% function renormalization means small physical down type Yukawa coupling
%% \begin{equation}
%%  y_{\bf 10} \sim y_* \quad {\rm while} \quad 
%%  y_{\bf 5^*} \sim \frac{y_*}{\sqrt{3(M_*L)^4}} \sim 0.1 y_* .
%% \end{equation}
This leads to a large wave function renormalization of the $\bar{H}^i$
because of the large volume, which may result in suppressed down type
Yukawa couplings $y_{\bf 5^*}$ compared with up type Yukawa couplings
$y_{\bf 10}$. 
% (rather small value of $\tan \beta \simeq 5 \sim 10$). 

%% consistency again 
We discussed the triangle anomaly cancellation on the orbifold
geometry. This is one of consistency conditions to be checked, 
but is not the only one.
There are further consistency conditions, and they are discussed in a
future publication\cite{future}.

Since the origins of the ${\bf Z}_4$ R symmetry are (local Lorentz and
$\U(1)_6$) gauge symmetries,
the low energy ${\bf Z}_4$ R symmetry is also gauged.
Thus, the discrete anomaly of the ${\bf Z}_4$ R symmetry 
must also be canceled out.
Fortunately, it is known that the ${\bf Z}_4$ R symmetry of the Table
\ref{tab:Z4-charge} has vanishing anomaly with a minimal extension of the 
present model as discussed in \cite{KMY}.

\section{Conclusions and Discussion}

In Ref.\cite{IWY} it has been shown that 
various features of the 
semi-simple unification model are simultaneously explained 
in a type IIB orientifold with a D3-D7 system. 
In this paper, we show that the ${\bf T}^6/{\bf Z}_{12}$ orientifold
model provides exactly the phenomenologically required matter content of 
the $\U(3)_{\rm H}$ sector (GUT-breaking sector)  
without unwanted light particles.
${\bf Z}_4$ R charge of each field determined from its property under the
rotation of the extra dimensional space can be the same as the desired one. 
Most of the superpotential relevant to the GUT breaking dynamics
and the missing partner mechanism are obtained from the interactions 
of the D3-D7 system.

Extension to the type IIB string theory gives us a geometrical 
interpretation of what is happening in this model.
Fist of all, the presence of the ``N=2 gauge interaction'' term
and the missing partner term becomes clear in the string theory. 
This is because both the
$X^{\alpha}_{(3)\;\beta}$ and the $\bar{H}_i=\Sigma^6_{(3)\;i}$ are 
fluctuations of the D3-branes and D7-branes, respectively, in the
D7-transverse directions, and expectation values of these fields ({\it
i.e.,} separation
between D3- and D7-branes) must provide masses to the D3-D7 open strings
$Q$ and $\bar{Q}$\cite{Polchinski}.
Secondly, the GUT breaking is regarded as a bound state formation
of the three D3-branes ($\U(3)_{\rm H}$) with five D7-branes
($\SU(5)_{\rm GUT}$). Three of the five
D7-branes form a bound state together with the three D3-branes, 
while two of them
keep the original nature, and that is how the triplet-double splitting
takes place. Indeed, the massless field $X^{\alpha}_{\;\;\beta}=
X^{\alpha}_{(3)\;\beta}$ acquires a mass with $\vev{X}=0$ after the GUT
breaking, which means that the D3-branes that could move freely in the 
D7-transverse directions are no longer able to leave the D7-brane
position \cite{bwb}.
Thirdly, there is a suggestion on the background geometry from 
the presence of the FI term. 
Recall that the type IIB string theory predicts an existence of twisted sector 
fields on orbifold fixed points. 
Discussion in \cite{DM} shows that there is a bilinear coupling 
between the $\sigma^3$-twisted sector fields and the $\U(1)_{\rm H}$ 
N=2 vector multiplet fields since 
tr($\gamma_{\sigma^3,3}$)$_{18 \times 18}$ $\neq 0$ (see
Eq.(\ref{eq:cp-3-s1-1}) and Eq.(\ref{eq:cp-3-s3-1})). 
In particular, the origin of the FI F term that induces 
the GUT breaking is interpreted as an expectation value of the 
$\sigma^3$-twisted NS-NS sector field. Under this interpretation, 
the FI F term indicates that the back ground geometry is not exactly 
the orbifold, but rather, some of the orbifold singularities 
are blown up and topologically nontrivial two cycles appear instead of 
the singularities. 
Finally, the $\U(3)_{\rm H}$ D3-branes at an ``N=2 fixed
point'' are 
regarded as fractional branes\cite{DGD}. They wrap the topologically 
nontrivial cycles discussed above and cannot move away from that place into 
D7-tangential directions, and that is how the unwanted NG modes are
eliminated.

%%  We constructed the model in ten dimensional supergravity and not in the
%%  string theory. String theory requires more stringent consistency
%%  conditions. Tadpole cancellation conditions that must be satisfied in
%%  orientifold compactification of any string theory are (probably) 
%%  not satisfied in this ${\bf T}^6/({\bf Z}_{12} \times {\bf
%%  Z}_2\vev{\Omega R_{89}})$ model.
%%  Furthermore, twisted sector matter content of the orbifold compactification 
%%  is not enough to explain the whole matter needed to be put by hand at
%%  the ${\bf Z}_{12}$ fixed point with only N=1 SUSY. And finally, 
%%  there are winding open strings in string theories, and we have to face
%%  with a six dimensional quadrangle/tetragon gauge anomaly of the 
%%  $\U(1)_{\rm H}$ 
%%  gauge symmetry after taking twice T-duality transformation 
%%  if the back ground geometry is exactly the orbifold limit 
%%  of the six dimensional torus.
%%  However, since the back ground geometry is not exactly in the orbifold
%%  limit of a torus, the semi-simple
%%  unification model constructed in the type IIB supergravity 
%%  do not necessarily contradict against consistency condition of the
%%  string / M theory.

The above points are very encouraging facts in the string theory.
However, even in the string theory it seems very difficult to obtain the 
three families of matter multiplets (${\bf 5}^* + {\bf 10}$) at the 
${\bf Z}_{12}$ fixed point.
This also strongly suggests that our manifold is not exactly the orbifold 
limit but rather it has more complicated structure around the fixed point.
If it is the case, it may be very difficult to derive further
consistency conditions originated from the string theory.
Nevertheless, we believe the connection of the present semi-simple
unification model in supergravity to the type IIB string theory to be
pursued in future investigations.

\section*{Acknowledgments}

The authors are grateful to Y.~Imamura for discussion and to M.~Kato for 
useful comments.
T.W. thank the Japan Society for the Promotion of Science for
financial support.
This work was partially supported by ``Priority Area: Supersymmetry and
Unified Theory of Elementary Particles (\# 707)'' (T.Y.).

\newpage

%
%%%%%%%%%%%%%%%%%%%%%%%%%%%%%%%%%%%%%%%%%%%%%%%%%%%%%%%%%%%%%%%
%
% NEW COMMANDS FOR THE BIBLIOGRAPHY
%
%%%%%%%%%%%%%%%%%%%%%%%%%%%%%%%%%%%%%%%%%%%%%%%%%%%%%%%%%%%%%%%
%\newcommand{\Journal}[4]{{\sl #1} {\bf #2} {(#3)} {#4}}
%\newcommand{\Journal}[4]{{#1} {\bf #2}, {#4} {(#3)}}
%                          journal, volume, year, page
\newcommand{\Journal}[4]{{\sl #1} {\bf #2} {(#4)} {#3}} %%% Elsevier-style
%\newcommand{\Journal}[4]{{#1} {\bf #2}, {#3} {(#4)}}   %%% PRD-style
%                          journal, volume, page, year 
%%%%%%%%%%%%%%%%%%%%%%%%%%%%%%%%%%%%%%%%%%%%%%%%%%%%%%%%%%%%%%%

\begin{table}
\begin{center}
\begin{tabular}{ccccccc}
\hline
Fields               & ${\bf 5}^*$,${\bf 10}$,${\bf 1}$ & $H$ , $\bar{H}$ &
                      $X^{\alpha}_{\;\beta} $ & $Q_i$,$\bar{Q}^i$ &
                                                $Q_6$  & $\bar{Q}^6$ \\
\hline
${\bf Z}_4$ R charge &                1               &     0           &
                             2                &      0            & 
                                                  2    &   -2        \\
\hline
\end{tabular}
\caption{Charge assignment of the ${\bf Z}_4$ R symmetry is given. ${\bf 1}$ denotes right-handed neutrino.}
\label{tab:Z4-charge}
\end{center}
\end{table}
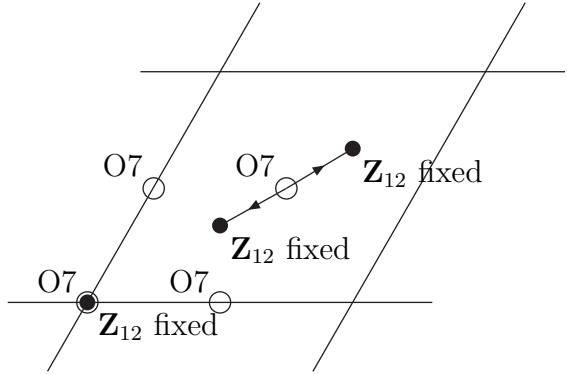
\begin{figure}
\begin{picture}(210,139)(-200,-50)
\Line(-30,0)(130,0) \Line(20,87)(180,87)
\Line(-15,-26)(65,113) \Line(85,-26)(165,113)
\Vertex(0,0){3} \Text(4,-4)[lt]{${\bf Z}_{12}$ fixed} 
\Vertex(50,29){3} \Text(54,25)[lt]{${\bf Z}_{12}$ fixed} 
\Vertex(100,58){3} \Text(104,54)[lt]{${\bf Z}_{12}$ fixed} 
\CArc(0,0)(4,0,360) \Text(-4,4)[rb]{O7}
\CArc(50,0)(4,0,360) \Text(46,4)[rb]{O7}
\CArc(25,43)(4,0,360) \Text(21,48)[rb]{O7}
\CArc(75,43)(4,0,360) \Text(71,48)[rb]{O7}
\ArrowLine(75,43)(100,58) \ArrowLine(75,43)(50,29)
\end{picture}
\caption{This figure shows a picture of the 3rd complex plane $z_3$ of the
${\bf T}^6/({\bf Z}_{12}\vev{\sigma} \times {\bf Z}_2\vev{\Omega R_{89}})$ 
 geometry. Circles are O7-plane positions and dots are ${\bf Z}_{12}
\vev{\sigma}$ fixed loci. We put six+six D7-branes for the 
$\SU(5)_{\rm GUT}$ on two dots without the circle. These two fixed loci 
 are a $\Omega R_{89}$-mirror pair and so are these six+six D7-branes.}
\label{fig:89}
\end{figure}
\end{document}